\documentclass[reprint,amsmath,amssymb,aps,prl,nofootinbib]{revtex4-1}

\usepackage{graphicx}
\usepackage{dcolumn}
\usepackage{bm}
\usepackage{hyperref}
\usepackage[usenames,dvipsnames]{color}

\hypersetup{
    colorlinks=true,         
    linkcolor=blue,          
    citecolor=red,        
    urlcolor=Violet             
}

\newcommand{\sfrac}[2]{\ensuremath{\textstyle{\frac{#1}{#2}}}}

\begin{document}

\preprint{APS/123-QED}

\title{Relationalism Evolves the Universe Through the Big Bang}

\author{Tim A. Koslowski$^1$, Flavio Mercati$^2$ and David Sloan$^3$}
\affiliation{$^1$ Instituto de Ciencias Nucleares, Universidad Nacional Aut\'onoma de M\'exico, Cd. Universitaria, 04510 Ciudad de M\'exico, D.F., M\'exico\\
$^2$ Perimeter Institute for Theoretical Physics, 31 Caroline Street North, Waterloo, ON N2L 2Y5, Canada\\
$^3$Beecroft Institute of Particle Astrophysics and Cosmology, Department of Physics, University of Oxford,
Denys Wilkinson Building, 1 Keble Road, Oxford OX1 3RH, UK}

\date{\today}

\begin{abstract}
\noindent 
All measurements are comparisons. The only physically accessible degrees of freedom (DOFs) are dimensionless ratios. The objective description of the universe as a whole thus predicts only how these ratios change collectively as one of them is changed. Here we develop a description for classical Bianchi IX cosmology implementing these relational principles. The objective evolution decouples from the volume and its expansion degree of freedom. We use the relational description to investigate both vacuum dominated and quiescent Bianchi IX cosmologies. In the vacuum dominated case the relational dynamical system predicts an infinite amount of change of the relational DOFs, in accordance with the well known chaotic behaviour of Bianchi IX. In the quiescent case the relational dynamical system evolves uniquely though the point where the decoupled scale DOFs predict the big bang/crunch. This is a non-trivial prediction of the relational description; the big bang/crunch is not the end of physics - it is instead a regular point of the relational evolution. Describing our solutions as spacetimes that satisfy Einstein's equations, we find that the relational dynamical system predicts two singular solutions of GR that are connected at the hypersurface of the singularity such that relational  DOFs are continuous and the orientation of the spatial frame is inverted.
\end{abstract}

\maketitle

The generic nature of singularities beyond which physics cannot be deterministically continued is a cornerstone of General Relativity (GR). The Hawking--Penrose theorems~\cite{Hawking} show that a large class of solutions of Einstein's equations are geodesically incomplete. In cosmological settings this leads to the big bang (or crunch) - the inevitable end of classical evolution. In this letter we show that when one considers cosmology from a relational perspective - constructing only observables available \textit{within} a toy model universe - this cornerstone is overturned. There exists a unique, deterministic, and entirely classical extension of Einstein's equations through the big bang/crunch. We achieve this result without appealing to quantum effects or new ad-hoc principles. Rather, the strict insistence on describing the dynamics in terms of relational variables alone ensures the existence and uniqueness of the evolution through the apparent singularity. The relational system predicts that the other side of the apparent singularity is a qualitatively similar yet quantitatively distinct cosmology \textit{inverted in spatial orientation.} 
We establish our result through a three-step process. First we rewrite the dynamics of a homogeneous (but not necessarily isotropic) cosmology entirely in terms of relational variables.  Second, we observe that the relational degrees of freedom form an autonomous subsystem that decouples from the evolution of the total size of the universe. Third, we show that the relational dynamical system remains deterministic while the system encounters (in finite physical time) a point at which Einstein's equations become singular. 


\section{Relational description of physics}

The description of the universe as a whole can not depend on external units of length or duration since all physical clocks and rods are part of the universe itself. The universe differs fundamentally from its subsystems in this aspect. The problem with GR's distance defining spacetime geometry is thrown into focus when considering cosmology. In a laboratory experiment an observer can easily justify the separation between the measuring apparatus (the external clock and rod) and the objects being measured. Cosmologists however are part of the universe and can not separate themselves form the studied system. Cosmological measurements are intrinsic, as rods and clocks are constructed from the dynamical objects in the universe. Units are constructed intrinsically using physical reference structures that define GR's notion of geometry. Hence, all dimensional quantities are intrinsic ratios. This leads us to concentrate on the dynamics of \textit{relational variables,} \textit{i.e.} dimensionless ratios and their relative infinitesimal variations\footnote{\textit{E.g.} the scale factor is \textit{not} a relational variable, while Misner's anisotropy parameters $\beta_+$, $\beta_-$~\cite{MTW} and the ratio of their variations $d\beta_+/d\beta_-$ are relational variables.}
\cite{Tutorial}. Remarkably, the dynamics can be expressed entirely in terms of relational variables, which turn out to evolve autonomously and predict all \textit{intrinsic observables} of GR.

In this letter we study spatially homogeneous cosmology (Bianchi IX) with a massless scalar field. This model is believed to correctly describe the near-singularity behaviour of full GR due to the BKL conjecture and Wheeler's insight that ``matter doesn't matter'' except a stiff component (such as a massless scalar field). In fact, the theorems of \cite{AR}, show that a dense set of inhomogeneous GR solutions obey the BKL conjecture, \textit{i.e.}  spatial points decouple in the approach to the singularity and evolve as independent Bianchi IX systems. Moreover, a massless scalar field is compatible with Standard Model physics\footnote{Interestingly, an RG improved gravitational action, as obtained in the functional renormalization group setting, also offers a mechanism to achieve this quiescent behavior~\cite{FrankGiulio}.} (\textit{e.g.} the Goldstone mode of the Higgs field). We thus distinguish two cases:\\
1. In absence of a massless scalar the approach to the singularity is given by the vacuum Bianchi IX evolution, in which the dynamics never actually reaches the singularity. It rather goes through an infinite amount of change, with infinitely many billiard-like `bounces' against steep triangular potential walls,  alternating with intervals of free geodesic evolution (Kasner epochs). This fact was observed by Misner~\cite{Misner_Mixmaster}, and its consequences for the status of the singularity was discussed in~\cite{MTW}. This has an important consequence in the relational framework, where physical clocks necessarily possess internal relational DOFs which register time.  An infinitesimal clock can not be treated as the idealized worldline of a point with its proper time, but has to be viewed as the infinitesimal limit of a sequence of ever smaller time-recording systems with internal structure  \cite{Ohanian}. It has been noted \cite{Ohanian} that the change of the internal relational DOFs  of an infinitesimal limit clock will be subject to the {\textit same} tidal effects as measured by its large counterparts. It follows that infinitesimal clocks register an unbounded lapse of time (\textit{i.e.}  change of internal relational DOFs), when the gravitational field experiences an infinite number of Kasner epochs.
It follows that the infinitesimal clocks, unlike their pointlike idealizations \textit{i.e.} proper time, will not reach the big bang/crunch in a finite time. \\
2. In the presence of a massless scalar one experiences ``quiescent" behaviour~\cite{Tutorial}. The potential becomes irrelevant for the dynamics and the equations of motion asymptote into a geodesic evolution. Matter clocks will measure a finite amount of change between the singularity and any other point.  It remains to investigate this case, because it is the one in which the singularity is reached in finite relational time and we have to establish what happens to the relational DOFs  there.

\section{Quiescent Bianchi IX cosmology}

\begin{figure}
\begin{center}
\includegraphics[width=0.4\textwidth]{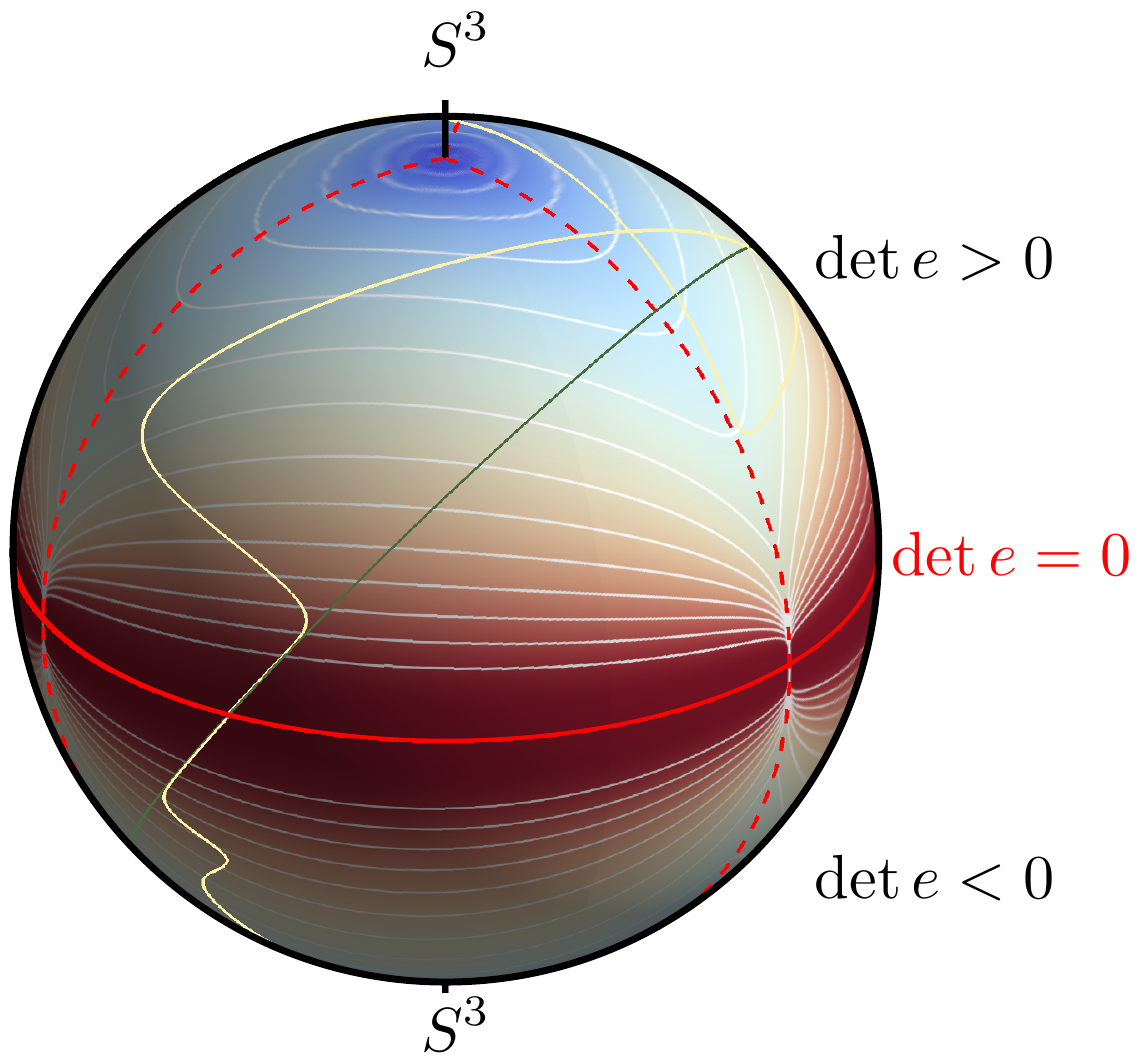}
\caption{\it
Representation of shape space: The poles $\beta=0,\pi$ represent isotropic geometries, the equator $\beta = \frac \pi 2$ degenerate ones. A typical solution is shown in yellow, with its asymptotic quiescent behaviour in green. 
The two hemisphere correspond to opposite spatial orientations. The shape potential $C(\alpha,\beta)$ is represented as a color plot on the sphere with equipotential lines in white and the GR singularity (the equator) in red. The FLRW spacetimes appear as unstable fixed points at the poles.}\label{CompactifiedShapeSpace}
\end{center}
\end{figure}

 Using triad variables, we describe Bianchi IX cosmology, \textit{i.e.}  homogeneous geometries on $S^3$, as:
\begin{equation}\label{4DLineElement}
ds^2 = -dt^2 + \delta_{ab}  e^a_i \, e^b_j \, dx^i dx^j \,.
\end{equation}
Imposing translational invariance, and fixing a global $SO(3)$ rotation, we write $e^a_i dx^i  = \pm v^{1/3} e^{\gamma_a} \sigma_a$, where $v$ is the spatial volume, $\gamma_a$ are three anisotropy parameters constrained by $\gamma_1 + \gamma_2 + \gamma_3 =0$, and $\sigma_a$ are the three translation-invariant one-forms on $S^3$. 
The $\pm$ in the definition of $e^a_i$ refers to the orientation of the spatial manifold and does not enter the metric. We can locally parametrize the anisotropy parameters with two Misner variables $q^1,q^2$, defined as: 
$\gamma_1 =  - q^1/\sqrt{6} - q^2/\sqrt{2}$,  $\gamma_2 =  - q^2/\sqrt{6} + q^1/\sqrt{2} $, $ \gamma_3 = \sqrt{\frac{2}{3}} q^2$ ,
 which coordinatize half of shape space. Useful global coordinates for shape space are the angles $(\alpha,\beta)$, defined by
{\medmuskip=0mu
\thickmuskip=0mu
\thinmuskip=0mu\begin{equation}\label{SphericalShapeCoordinates}
\left( \begin{array}{c}q^1\\q^2\end{array}\right) =|\tan\beta| \left( \begin{array}{c}
\cos\alpha \\ \sin \alpha
\end{array}\right), \, ~~~  \textrm{sign} (\det e) = \textrm{sign} (\tan \beta) \,,
\end{equation}}
where the sign of $\beta$ represents the two possible orientations of $e^a_i$. $(\alpha,\beta)$ are spherical coordinates for the representation of shape space shown in  Fig.~\ref{CompactifiedShapeSpace}.

We denote the shape momenta canonically conjugate to $q^a$  by $p_a$, the variable conjugate to $v$ by $\tau$, the (homogeneous) massless scalar field by $\varphi$ and its momentum density by $\pi$. Due to dynamical similarity~\cite{ArrowPaper,EntropyPaper}, only the latter will appear in the equations. Einstein's equations are generated by the ADM Hamiltonian~\cite{ADM} (which is constrained to vanish)
\begin{equation}\label{BIX-Hamiltonian}
 H=p_1^2+p_2^2+\frac{\pi^2}{2}-\frac 3 8\,\tau^2\,v^2-v^\frac{4}{3}\,C(q^1,q^2)\approx 0,
\end{equation}
where $C(q^1,q^2)$,  the shape potential shown in Fig.~\ref{CompactifiedShapeSpace}, is
{\medmuskip=0mu
\thinmuskip=0mu
\thickmuskip=0mu
\begin{equation}\label{ShapePotential}
\begin{aligned}
C(q^1,q^2) &= F(2\,q^2)+F(q^1\sqrt{3}-q^2)+F(-q^1\sqrt{3}-q^2) \,,
\\
F(x) &= e^{- x/\sqrt{6}} - {\sfrac 1 2} e^{2 x/\sqrt{6}} \,.
\end{aligned}
\end{equation}}
The equations of motion [using a vector notation $\vec q = (q^1 , q^2)$, $\vec p = (p_1 , p_2)$] are
\begin{equation}\label{BIXeqOfMotion}
\begin{aligned}
&\dot{\vec q} = 2 \, \vec p  \,,&  &\dot{\vec p} =  v^{4/3} \, \vec{\nabla} C(\vec q) \,,&  
\\
&\dot  v = - {\sfrac 3 4} v^2 \tau \,,&  &\dot  \tau = {\sfrac 4 3} v^{1/3} C(\vec q) + {\sfrac 3 4}  v \tau^2 \,.&
\end{aligned}
\end{equation}
where $\vec{\nabla} C= \left( \frac{\partial  C}{\partial {q^1}} , \frac{\partial  C}{\partial {q^2}}\right)$, and `` $\dot{}$ '' is the derivative w.r.t. the coordinate time $t$ appearing in~(\ref{4DLineElement}). We will now consider the relational description of this system and investigate the existence and uniqueness of solutions using the Picard-Lindel\"of theorem, which states that a system of differential equations $y_a'(x)=F_a(x,y)$ possesses a unique solution for the initial value problem $y_a=y_a^o$ at $x=x_o$ if $F_a$ is continuous in $x$ and Lipschitz-continuous in $y_a$ in a neighbourhood of $x_o$. We note that this is a coordinate-dependent statement, because the question whether the $f_a$ are continuous and Lipschitz depends on which variables $y_a$ one uses. For example the systems $F=(y_2,y_3,y_2/f(x))$ and $F=(u_2,u_3e^{\int^x_{x_o}\frac{ds}{f(s)}},0)$ are equivalent when $f(x)\ne 0$. However, considering $f(x)=x$ reveals that at $x_o=0$ the former system fails, while the latter satisfies the conditions of the Picard-Lindel\"of theorem.

\noindent\textbf{Decoupling of scale.}

With the variables at our disposal, we can form the following three independent dimensionless and scale-invariant combinations:
\begin{equation}\label{BIX_DefXiSigmaKappa}
\xi = \frac{|\pi|}{p} \,,
\qquad
\sigma = \frac{v \, |\tau|}{p} \,,
\qquad
\kappa = \frac{v^{2/3}}{p} \,.
\end{equation}
The Hamiltonian constraint~(\ref{BIX-Hamiltonian}) in those variables reads
\begin{equation}\label{Dimensionless_BIX_Ham_const}
\mathcal H = p^2 \left[ {\sfrac 3 8} \, \sigma^2 -   \left(1  + {\sfrac 1 2} \xi^2
 \right) +  \kappa^2 \, C(q^1,q^2) \right] \approx 0 \,,
\end{equation}
which, for dynamically nontrivial solutions in which $p\neq 0$ (this excludes FRLW, in which there is no shape evolution) implies a relationship between $\sigma$, $\xi$, $\kappa$, $q^1$ and $q^2$.
It is easy to show that in quiescent solutions, which reach the equator of shape space $\beta = \frac \pi 2$, the factor $\kappa^2 \, C(\vec q) $ vanishes as  $\beta \to \frac \pi 2$, and therefore:
\begin{equation}
1  + {\sfrac 1 2} \xi^2 - {\sfrac 3 8}  \sigma^2 \xrightarrow[\beta \to \frac \pi 2]{} 0 \,,
\end{equation}
so the variables $\xi$, $\sigma$ and $\kappa$ are asymptotically redundant and they do not provide a good parametrization of phase space near the equator. We need to replace one of the variables with something that takes a generic value at $\beta = \frac \pi 2$. A good choice is:
\begin{equation}
\omega =  \text{sign} (\tan \beta) \left[ \frac{\vec q \cdot \vec p}{p} - \frac{2 p \log \left( v^2 \tau^6\right)}{3\,v \tau} \right] \,,
\end{equation}
which is an asymptotically conserved quantity (it is preserved by the Bianchi I equations of motion), and moreover, thanks to the sign of $\beta$ factor, it is continuous through the singularity in Bianchi I solutions.

We can now express the equations of motion~(\ref{BIXeqOfMotion}) in terms of $\omega$, $\sigma$, $\xi$ and the angular coordinates on the shape sphere [$\alpha = \arctan(q^2/q^1)$,  $\beta = (\sfrac{1 - s}{2}) \pi  + s  \arctan \sqrt{(q^1)^2 + (q^2)^2}$, where $s= \text{sign} (\tan \beta)$], plus the scale-free `angular momentum' variable:
\begin{equation}
\gamma = \frac{q^1 p_2 - q^2 p_1}{p} \,,
\end{equation}
which is conserved by the quiescent/Bianchi I evolution, and therefore is conserved at the equator.
Moreover we can parametrize the equations with the arc-length on shape space $(dq^1)^2 + (dq^2)^2 = d\ell^2$, to obtain:
{\thinmuskip=0mu
\medmuskip=0mu
\thickmuskip=0mu
\begin{equation}\label{BIX_intrinsiceq_2}
\begin{aligned}
\alpha' ~=& ~ \gamma ~ \cot^2 \beta \,,
\qquad
\beta' ~=~ \cos^2 \beta ~ \sqrt{1 - \frac{\gamma^2}{\tan^2 \beta}} \,,
\\
\gamma' ~=&~ f_\gamma ~\epsilon \,,
\qquad
\omega' ~=~ f_\omega ~ \epsilon  \,,
\qquad
\sigma' ~= f_\sigma ~ \epsilon \,.
\end{aligned}
\end{equation}}
where $x' = \frac{d x}{d \ell}$ and
\begin{equation}
\epsilon = e^{{\frac \sigma 2} \left(\omega - \sqrt{\tan^2 \beta - \gamma^2}  \right) } \,,
\end{equation}
and where $f_\gamma$, $f_\omega$ and $f_\sigma$ are functions of $\alpha$, $\beta$, $\gamma$, $\omega$ and $\sigma$ (their definition is in appendix). The nontrivial fact of this reformulation is that the variable $p$ completely decouples from the equations of motion, and it is not necessary anymore to determine the solution curve on shape space.

A straightforward application of the Picard-Lindel\"of theorem implies that this  dynamical system possesses a unique solution for any initial values  $\left(\alpha,\beta \neq {\frac \pi 2},\gamma,\omega,\sigma>0 \right)$. This allows us to conclude that the relational description of the system is predictive everywhere, except possibly at the equator of shape space.

\noindent\textbf{Ephemeris equations.}

Equations~(\ref{BIX_intrinsiceq_2}) do not contain any information regarding scale or duration. Units of scale and time need to be fixed once and for all at a point on a solution, and they are completely immaterial. Their subsequent evolution is entirely determined by the shape degrees of freedom, and it can be calculated using two `ephemeris' equations  (they are just the equations of motion of $p$ and $v$ in arclength parametrization):
\begin{equation}
\begin{aligned}
\frac{d \log p}{d \ell} =& \frac{e^{{\frac \sigma 2} \left(\omega - \sqrt{\tan^2 \beta - \gamma^2}  \right) }}{2\tan^2 \beta} \left(  \gamma ~ \frac{\partial C}{\partial \alpha}  \right.
\\
&\left.
 + s ~\sin (2 \beta) \sqrt{\tan^2 \beta - \gamma^2}  \frac{\partial C}{\partial \beta}  \right) \,.
\\
\frac{d \log v}{d \ell} =& - \frac 3 8 ~ s ~ \sigma \,,
\end{aligned}
\end{equation}
What makes these `ephemeris' equations is the fact that the unknown variables $p$ and $v$ do not appear on the right-hand-side, so they `piggy-back' on the evolution of the shape variables. Moreover the equations only determine the logarithms of $p$ and $v$, and therefore their solution are defined modulo a constant rescaling: this is the arbitrariness in fixing units at one point on the solution.

To reach the singularity $v=0$ from any finite point in shape space, either the rhs of the ephemeris scale equation diverges  (which requires extra symmetry) or an infinite distance of kinematic arc-length is reached. The second condition is generic and states that the singularity is reached whenever the curve in shape space reaches the equator, since each point on the equator has infinite kinematic arc-length distance from any other point. We thus find the singularity condition $ \beta(\ell) =\frac{\pi}{2}$, where, in the spacetime description, the big bang/crunch occurs.

\section{Continuation through the singularity}

We now study the well-posedness of the system~(\ref{BIX_intrinsiceq_2}) in the only remaining region: the equator of shape space $\beta = \frac \pi 2$. For this purpose we use a different intrinsic parametrization, in which the parameter stays finite at the equator (the arc-length $d\ell$ diverges there): the $\beta$ coordinate. To do so, it is sufficient to divide Eqs.~(\ref{BIX_intrinsiceq_2}) by $\beta' = \cos^2 \beta ~ \sqrt{1 - \frac{\gamma^2}{\tan^2 \beta}}$. Unlike the arc-length, this parametrization is not good everywhere on the solution curve, because $\beta$ is not monotonic, but it is a good parametrization in a neighbourhood of the equator, where $\beta' \sim (\beta - \frac{\pi}{2})^2 + \mathcal{O} (\beta - \frac{\pi}{2})^4 \geq 0$. The new equations read
{\thinmuskip=0mu
\medmuskip=0mu
\thickmuskip=0mu\begin{equation}\label{BIX_intrinsiceq_betaparametrization}
\begin{aligned}
\frac{d \alpha}{d \beta}  =  \frac{\gamma}{\sin^2 \beta ~ \sqrt{1 - \frac{\gamma^2}{\tan^2 \beta}}}  \,,
~~
\frac{d \gamma}{d \beta} = \frac{f_\gamma ~ \epsilon }{ \cos^2 \beta ~ \sqrt{1 - \frac{\gamma^2}{\tan^2 \beta}}}   \,,
\\
\frac{d \omega}{d \beta} = \frac{f_\omega ~ \epsilon}{ \cos^2 \beta ~ \sqrt{1 - \frac{\gamma^2}{\tan^2 \beta}}} \,, ~~
\frac{d \sigma}{d \beta} = \frac{f_\sigma ~ \epsilon}{ \cos^2 \beta ~ \sqrt{1 - \frac{\gamma^2}{\tan^2 \beta}}} \,.
\end{aligned}
\end{equation}}
It is easy to prove that $\frac{f_i ~ \epsilon}{ \cos^2 \beta ~ \sqrt{1 - \frac{\gamma^2}{\tan^2 \beta}}} $ is Lipschitz continuous around  $\beta = {\frac \pi 2}$ (and in particular they all vanish there), if the following bound is satisfied:
\begin{equation}\label{BoundOnSigma}
\begin{aligned}
\sigma >  \sqrt{\sfrac 8 3} \max_{\alpha\in(0,2\pi]} \left( 2 \sin \alpha, \pm \sqrt{3} \cos \alpha -  \sin \alpha  \right) \,.
\end{aligned}
\end{equation}
This `quiescence' bound on $\sigma$ comes from the requirement that the quantities $\epsilon ~ C(\alpha,\beta)$, $\epsilon ~ \frac{\partial C}{\partial \alpha}$ and $\epsilon ~\frac{\partial C}{\partial \beta}$ 
tend to zero as $\beta \to \frac \pi 2^\pm$.
We thus conclude from the application of the Picard-Lindel\"of theorem that the equations of motion are deterministic through the point $\beta=\pi/2$.

\noindent\textbf{Quiescent/Bianchi I behaviour}

Near the equator, the potential terms in~(\ref{BIX_intrinsiceq_betaparametrization}) can be neglected. In this limit, the equations turn into Bianchi I equations (Kasner regime), whose solutions are straight lines in the $q^i$ plane, and the variables $\gamma$, $\omega$ and $\sigma$ become conserved:
\begin{equation}\label{BIeqs1}
\frac{d \alpha}{d \beta}  =  \frac{\gamma}{\sin^2 \beta \, \sqrt{1 - \frac{\gamma^2}{\tan^2 \beta}}}  \,, ~~ \frac{d \gamma}{d \beta} = \frac{d \omega}{d \beta} = \frac{d \sigma}{d \beta} = 0 \,.
\end{equation}
The general solution to the above equation is  $\alpha = \arcsin \left( \frac{- s \, \gamma}{\tan \beta} \right) + \text{\it{const.}}$, $\gamma =  \text{\it{const.}}$, $\omega =  \text{\it{const.}}$, $\sigma =  \text{\it{const.}} $,  which represents a great circle on the shape sphere, evolving smoothly through the equator. 
This is how we confirm the well known fact that quiescent Bianchi IX system asymptotes into into Bianchi I behaviour. However, the relational system does more: It contains two additional degrees of freedom: $\sigma$ and $\omega$, whose equations of motion stay well defined at the equator. This means that the relational system evolves through the equator, where General Relativity would place the big bang/crunch.

This establishes the main technical result of this paper.
The interpretation of this result in terms of spacetime geometry is that the purely relational description of quiescent Bianchi IX glues two quiescent Bianchi IX spacetimes at the big bang in such a way that the relational variables  $(\alpha,\gamma,\omega,\sigma)$ are continuous. Recall also that the crossing of $\beta=\frac{\pi}{2}$ implies an inversion of spatial orientation.

Notice that singularities can only happen at the equator $\beta=\frac \pi 2$, or in the exceptional FLRW case. The latter is represented on shape space by an unstable fixed point at a pole $\beta = 0,\pi$, so it is dynamically unattainable. The points on the equator represent degenerate geometries in which some ratios between anisotropy parameters diverge. Furthermore, all quiescent solutions satisfy (\ref{BoundOnSigma}) when approaching the equator. This means that any initial condition set at $\beta=\frac \pi 2$ is required to satisfy (\ref{BoundOnSigma}).

\noindent\textbf{Generic nature of the result.}
Our discussion so far only applies to \textit{homogeneous} cosmology. However, it was shown in \cite{AR} that there is a dense set of \textit{inhomogeneous} solutions of GR that satisfy the BKL conjecture, \textit{i.e.}  as the singularity approaches each spatially-separated point decouples and evolves like an autonomous Bianchi IX system. This allows us to extend our result immediately to this dense set. As yet it remains to be seen if this dense set is the complete set of all solutions to GR.

\section{Discussion}

We showed that the relational description of quiescent Bianchi IX universes evolves through the big bang, which is not a singularity of the equations on shape space. We will therefore call it a `Janus point', because it is the point of qualitative time-symmetry of each solution~\cite{ArrowPaper,EntropyPaper,JanusPointAnswerToZeh}. The Janus-point data, \textit{i.e.}  the specification of   $(\alpha,\gamma,\omega,\sigma)$ at $\beta=\frac{\pi}{2}$, determines a unique curve on the two hemispheres of shape space (with a single intersection with the equator) that can be effectively described as two quiescent Bianchi IX spacetimes glued together at the big bang, where a change of orientation occurs. The prediction of a classical change of orientation of the spatial manifold at the big bang could have profound implications for discrete symmetries in particle physics, particularly regarding matter/antimatter asymmetry.

In vacuum dominated Bianchi IX (\textit{i.e.}  in absence of massless scalars) on the other hand, one finds that physical time (measured by the change of shape of any finite clock) will go on forever in the relational description of vacuum-dominated Bianchi IX cosmology. These curves do not terminate in shape space. As such an intrinsic observer will never encounter any singularity. 

We established that big bang/crunch singularities in homogeneous cosmologies with compact topology are spacetime artifacts, that do not have any physical (\textit{i.e.}  completely relational) meaning. This has important consequences for a dense set of GR solutions (those described in~\cite{AR}) whose behaviour near the cosmological singularity is completely described by an independent Bianchi IX universe at each point. Moreover, the BKL conjecture posits that this behaviour is generic in GR. Finally, this result allows discussing the typicality of universes in terms of their Janus point data, and the spontaneous emergence of an arrow of time (along the lines of \cite{ArrowPaper,EntropyPaper,JanusPointAnswerToZeh}).

\section*{Acknowledgments}

\begin{acknowledgments}
The authors are indebted to Julian Barbour for his hospitality and encouragement during the development of this project, and his recognition of the existence and significance of Janus points. Our work profited from regular intensive discussions with him, many at his home College Farm. This work was to a great extent funded by the Foundational Questions Institute (FQXi). The authors benefitted from useful conversations and suggestions from Bei-Lok Hu, Steve Carlip, Henrique Gomes, Sean Gryb and Niall \'O Murchadha. Perimeter Institute is supported by the Government of Canada through Industry Canada and by the Province of Ontario through the Ministry of Economic Development and Innovation. This research was also partly supported by grants from the John Templeton Foundation.
\end{acknowledgments}

\appendix
\section{Appendix}
\label{appendix}

The intrinsic equations (\ref{BIX_intrinsiceq_2}) and (\ref{BIX_intrinsiceq_betaparametrization}) are defined in terms of the functions $f_\gamma$, $f_\omega$ and $f_\sigma$ which are functions only of the intrinsic variables. These are:

{\thinmuskip=0mu
\medmuskip=0mu
\thickmuskip=0mu
\begin{equation}
\begin{aligned}
f_\gamma =&\textstyle \frac{\sigma^2}{2}  \sqrt{1 -\frac{\gamma^2}{\tan^2 \beta}} \left( \sqrt{1 -  \frac{\gamma^2}{\tan^2 \beta} }  \, \frac{\partial C}{\partial \alpha} - \gamma \cos^2 \beta \,  \frac{\partial C}{\partial \beta}   \right) \,,
\\
f_\omega =&\textstyle {\frac 2 3} s ~ \left[ \sigma \left(\sqrt{\tan^2 \beta - \gamma^2} - \omega \right) - 4 \right] C(\alpha,\beta) \\
&\textstyle+\frac{\sigma^2}{2} |\cos \beta|^3 \left[    \left( 2  \gamma^2 - \tan^2 \beta \right) +  \omega \sqrt{\tan^2 \beta - \gamma^2  } \right]  \frac{\partial C}{\partial \beta}
\\
&\textstyle+ \frac{\gamma ~\sigma^2}{2 \, \tan^2 \beta} \left( 2 \sqrt{\tan^2 \beta - \gamma^2}  - \omega \right)  \frac{\partial C}{\partial \alpha}  \bigg{\}} \,,
\\
f_\sigma =&\textstyle{\frac 2 3} s \, \sigma^2 ~C(\alpha,\beta)
\\
&\textstyle-\frac{\sigma^3}{2 \, \tan\beta}  \left(  \frac{\gamma}{\tan\beta} ~ \frac{\partial C}{\partial \alpha} 
+  s ~ \cos^2\beta \sqrt{\tan^2 \beta - \gamma^2} \frac{\partial C}{\partial \beta}\right) \,.
\end{aligned}
\end{equation}}

Note that these functions are defined in terms of the shape potential $C$ and its derivatives, and with the appropriate choices of $C$ hold for all class A Bianchi models.


\begin{thebibliography}{10}

\bibitem{Hawking}
S.~W. Hawking and R.~Penrose, ``{The Singularities of gravitational collapse
  and cosmology},''
\href{http://dx.doi.org/10.1098/rspa.1970.0021}{{\em Proc. Roy. Soc. Lond.}
  {\bfseries A314} (1970) 529--548}.

\bibitem{AR}
L.~Andersson and A.~D. Rendall, ``{Quiescent cosmological singularities},''
  {\em Commun. Math. Phys.} {\bfseries 218} (2001) 479--511.

\bibitem{MTW}
C.~W. Misner, K.~S. Thorne, and J.~A. Wheeler, {\em Gravitation}.
\newblock Macmillan, 1973.

\bibitem{Tutorial}
F.~Mercati, {\em Shape Dynamics: Relativity and Relationalism}.
\newblock Oxford Univ. Press, 2016.
\newblock \href{http://arxiv.org/abs/1409.0105}{{\ttfamily arXiv:1409.0105}}.
\newblock (preliminary version on the arXiv).

\bibitem{FrankGiulio}
G.~D'Odorico and F.~Saueressig, ``Quantum phase transitions in the
  Belinsky-Khalatnikov-Lifshitz universe,''
  \href{http://dx.doi.org/10.1103/PhysRevD.92.124068}{{\em Phys. Rev.}
  {\bfseries D92} no.~12, (2015) 124068},
  \href{http://arxiv.org/abs/1511.00247}{{\ttfamily gr-qc:1511.00247}}.

\bibitem{Misner_Mixmaster}
C.~W. Misner, ``Mixmaster Universe,''
  \href{http://dx.doi.org/10.1103/PhysRevLett.22.1071}{{\em Phys. Rev. Lett.}
  {\bfseries 22} no.~20, (1969) 1071--1074}.

\bibitem{Ohanian}
H.~C. Ohanian, ``What is the Principle of Equivalence?,'' {\em Am. J. Phys.}
  {\bfseries 45} no.~10, (1977) 903--909.

\bibitem{ArrowPaper}
J.~Barbour, T.~Koslowski, and F.~Mercati, ``{Identification of a gravitational
  arrow of time},''
  \href{http://dx.doi.org/10.1103/PhysRevLett.113.181101}{{\em Phys. Rev.
  Lett.} {\bfseries 113} no.~18, (2014) 181101},
\href{http://arxiv.org/abs/1409.0917}{{\ttfamily arXiv:1409.0917 [gr-qc]}}.

\bibitem{EntropyPaper}
J.~Barbour, T.~Koslowski, and F.~Mercati, ``{Entropy and the Typicality of
  Universes},''
\href{http://arxiv.org/abs/1507.06498}{{\ttfamily arXiv:1507.06498 [gr-qc]}}.

\bibitem{ADM}
R.~L. Arnowitt, S.~Deser, and C.~W. Misner, ``The dynamics of general
  relativity,'' \href{http://arxiv.org/abs/gr-qc/0405109}{{\ttfamily
  arXiv:gr-qc/0405109}}. In Gravitation: an introduction to current research,
  Louis Witten ed., chapter 7, pp 227--265.

\bibitem{teschl2012ordinary}
G.~Teschl, {\em Ordinary differential equations and dynamical systems},
  vol.~140.
\newblock American Mathematical Soc., 2012.

\bibitem{JanusPointAnswerToZeh}
J.~Barbour, T.~Koslowski, and F.~Mercati, ``Janus Points and Arrows of Time,''
  \href{http://arxiv.org/abs/1604.03956}{{\ttfamily arXiv:1604.03956 [gr-qc]}}.

\end{thebibliography}

\end{document}